\documentclass[reprint,twocolumn,amsmath,showpacs,superscriptaddress,amssymb,aps,prb]{revtex4-1}

\usepackage{graphicx}
\usepackage{graphics}
\usepackage{amsmath}
\usepackage{dcolumn}
\usepackage{bm}

\begin{document}


\title{Ferroelectricity-induced asymmetrical two-dimensional electron gas in superlattices consisteing of insulating GdTiO$_3$ and ferroelectric BaTiO$_3$}

\author{Xue-Jing Zhang}
\affiliation{Beijing National Laboratory for Condensed Matter Physics, Institute of Physics, Chinese Academy of Sciences, Beijing 100190, China}
\author{Bang-Gui Liu}\email{bgliu@iphy.ac.cn}
\affiliation{Beijing National Laboratory for Condensed Matter Physics, Institute of Physics, Chinese Academy of Sciences, Beijing 100190, China}
\affiliation{School of Physical Sciences, University of Chinese Academy of Sciences, Beijing 100190, China}

\date{\today}

\begin{abstract}
Two-dimensional electron gas due to semiconductor interfaces can have high mobility and exhibits superconductivity, magnetism, and other exotic properties that are unexpected in constituent bulk materials. We study crystal structures, electronic states, and magnetism of short-period (BTO)$_m$/(GTO)$_2$ ($m$=2 and 4) superlattices consisting of ferroelectric BaTiO$_3$ (BTO) and ferrimagnetic insulating polar GdTiO$_3$ (GTO) by first principles calculations. Our investigation shows that the middle Ti-O monolayer in the GTO layer becomes metallic because the ferroelectricity in the insulating BTO layer induces an inhomogeneous electric field against the polarity-produced electric field in the GTO layer and thus differentially changes the d energy levels of the three Ti-O monolayers related with the GTO layer. Through avoiding electron reconstruction, the ferroelectric polarization also makes the electronic states and magnetism of two interfacial Ti-O monolayers become substantially different from those in the GTO/SrTiO$_3$ superlattices without ferroelectricity. Such superlattices are interesting for potential spintronics applications because of their unique asymmetrical two-dimensional electron-gas properties and possible useful spin-orbit effects.
\end{abstract}

\maketitle


\section{Introduction}

Perovskite heterostructures often exhibit unusual physical properties that are absent in the constituent bulk materials \cite{a1,a2}. It is well known that two-dimensional electron gases (2DEG) can be formed at LaAlO$_3$/SrTiO$_3$ interfaces and such 2DEG can have high mobility and exhibit superconductivity, magnetism, and other exotic properties that are unexpected in the corresponding bulk materials \cite{a3,a4,a5}.
2DEG at interfaces between Mott insulators and band insulators have attracted significant attention. Stoichiometric GdTiO$_3$ (GTO) is a Mott-Hubbard insulator that has a gap of 1.8 eV in terms of photoluminescence measurements.\cite{1} In bulk GTO, the ferromagnetic (FM) Ti array is antiparallel to the FM Gd array, resulting in a net ferrimagnetism with Curie temperature TC of 32 K.\cite{2,3,4} Compressive strain can make the ferromagnetic ordering of the Ti atoms change into a G-type antiferromagnetic (AFM) ordering on the LaAlO$_3$ substrate and A-type AFM ordering on the YAlO$_3$ substrate.\cite{5} The transport measurements of the 2DEGs at SrTiO$_3$(STO)/LaTiO$_3$(LTO) and STO/GTO interfaces exhibit an interfacial density values with the order of $3\times 10^{14}$ cm$^{-2}$, approximately 1/2 electron per surface unit cell, which are essentially those predicted by electronic reconstruction.\cite{8,9,10,11} This electronic reconstruction implies that when a LTO (or GTO) is combined with STO, the Ti cation in the TiO$_2$ atomic layers between LaO (or GdO) and SrO monolayers actually has an averaged valence of 3d$^{0.5}$.\cite{12,13,14} A charge modulation of over $10^{14}$ cm$^{-2}$ electrons in 2DEGs formed in STO/GTO heterostructure field-effect transistors has been reported.\cite{15,16} Recently, it has been shown, through combining experiment and theory, that in short-period STO/GTO superlattices, the GTO layer is insulating and non-polar, and conduction electrons are located in the STO layer, at neither of the two interfacial Ti-O monolayers.\cite{11}

In order to achieve better 2DEG, however, we use ferroelectric BaTiO$_3$ (BTO)\cite{6,7} instead of STO to construct our (BTO)$_m$/(GTO)$_2$ superlattices, and then study the crystal structures and electronic properties of the superlattices with $m$ = 2 and 4 because the BTO ferroelectricity can survive in high-quality ultra-thin BTO layers\cite{30a}. Our calculation and analysis show that both the ferroelectricity in the BTO layer and the polarity in the GTO layer survive through helping each other. The interfacial properties are substantially influenced by the ferroelectricity, charge polarity, and electronic correlations. We found that charge carriers are located not in the ferroelectric BTO layer, but in the two Ti-O monolayers related with the polar GTO layer. In the absence of electron reconstruction, the two interfacial Ti-O monolayers have very different magnetic moments and conductive properties. These imply that the 2DEG can be manipulated through controlling the direction of the ferroelectric polarization. Therefore, such superlattices are very interesting for designing switchable functional devices. More detailed results will be presented in the following.

\section{Computational details}

Our first-principles calculations are performed using the projector-augmented wave method within the density-functional theory\cite{dft1,dft2}, as implemented in the Vienna Ab-initio Simulation Package (VASP).\cite{17,18} The plane wave energy cutoff is set to 600 eV. For the exchange-correlation potential, we use the generalized gradient approximation (GGA) by Perdew, Burke, and Ernzerhof.\cite{19} The rotationally invariant GGA+U method is employed with $U=5.00$ and $J=0.64$ eV for Ti 3d and $U=7.70$ and $J=0.70$ eV for Gd 4f electrons.\cite{20,21} The cell volume is relaxed and the ionic positions are optimized using a $\Gamma$-centered $6\times 6\times 1$ k-grid. The electronic structure calculations were performed by using a $\Gamma$-centered $12\times 12\times 1$ k-grid. Our convergence standard requires that the Hellmann-Feynmann force on each atom is less than 0.01 eV/\AA. The convergence standard for the total energy is chosen to be $10^{-5}$ eV.

\section{Results and discussion}

\subsection{Optimized structures}

Band insulator BTO exhibits room-temperature ferroelectric phase with a semiconductor gap of 3.27 eV. \cite{6,7} In the GTO (001) direction, Gd$^{3+}$O$^{2-}$ and Ti$^{3+}$O$_2^{4-}$ monolayers appear alternately, carrying formal +1 and -1 charges, respectively, but in the non-polar BTO (001) direction both Ba$^{2+}$O$^{2-}$ and Ti$^{4+}$O$_2^{4-}$ monolayers are charge neutral. Moreover, both GTO and BTO share a common constituent, the TiO$_2$ atomic layer (Ti-O monolayer). The paraelectric cubic BTO has space group Pm3m with experimental lattice constant of 4.00 \AA{} and the BTO (001) layer has  $\sqrt{2}\times\sqrt{2}$ periodicity with experimental lattice constant $\sqrt{2} a$ of 5.65 \AA.\cite{22} Bulk GTO has an orthorhombic Pbnm structure with experimental lattice constants of $a=5.39$ \AA, $b=5.69$ \AA, and $c=7.66$ \AA{}, and the GTO (001) layer has $1\times 1$ periodicity.\cite{23} Thus, the mismatch amounts to 4.71\% and 0.71\% for the $a$ and $b$ axes. The optimized lattice constants are $a=4.06$ \AA, with $\sqrt{2} a$ of 5.74 \AA, for BTO, and $a=5.46$ \AA, $b=5.79$ \AA, and $c=7.81$ \AA{} for GTO, consistent with previous calculation values.\cite{1}

\begin{figure}[!htbp]
\centering  
\includegraphics[clip, width=8.5cm]{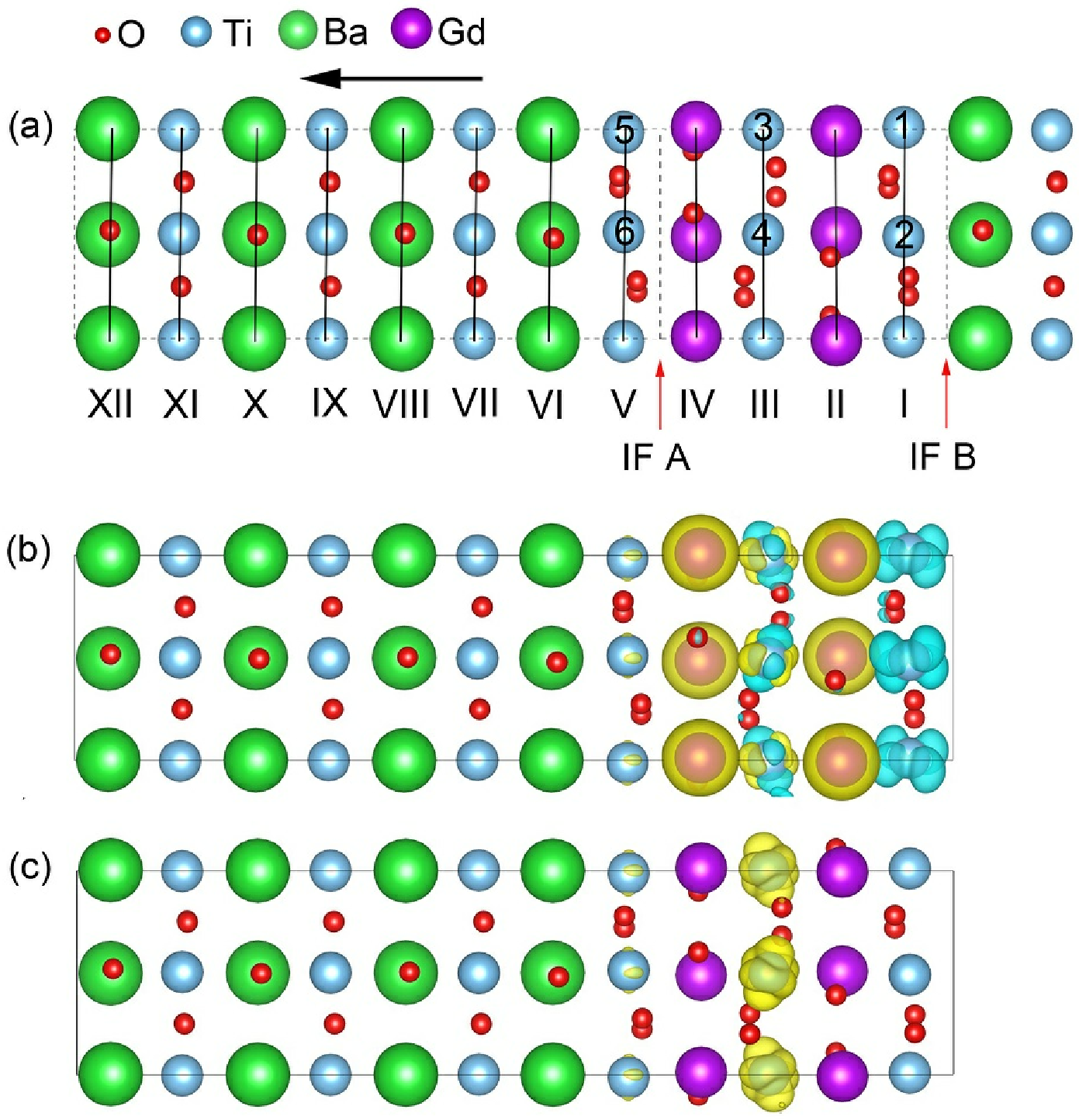}
\caption{(Color online) (a) Side view of the structures of (BTO)$_4$/(GTO)$_2$ superlattice. (b) Spin density for the (BTO)$_4$/(GTO)$_2$ superlattices, The yellow and cyan isosurfaces ($\pm 0.008|e|$/\AA$^3$) represent positive and negative spin density, respectively. (c) The yellow isosurfaces (0.004$|e|$/\AA$^3$) represent charge density ranging from -0.5 to 0 eV in (BTO)$_4$/(GTO)$_2$ superlattice.}\label{edge}
\end{figure}

The optimized structure of (BTO)$_4$/(GTO)$_2$ is shown in Fig. 1(a). The optimized lattice constants are $a=5.67$ \AA{} and $b=5.74$ \AA{} for (BTO)$_4$/(GTO)$_2$, meaning a compressive strain of 1.23\% in the $a$ axis for the BTO layer and a tensile strain of 3.77\% for the GTO layer. There is a zero strain in the $b$ axis for the BTO layer and a tensile strain of 0.87\% for the GTO layer. The calculated band gap of bulk GTO is 2.27 eV, consistent with previous calculation values\cite{1} and slightly greater than that of 2.02 eV calculated by a hybrid functional.\cite{24} The optimized result for the (BTO)$_2$/(GTO)$_2$ superlattice turns out to be very similar to that of the (BTO)$_4$/(GTO)$_2$ superlattice and hence we shall not present its detail. The magnetic moment value of each atom in the (BTO)$_4$/(GTO)$_2$ is visualized in Fig. 1(b). It can be seen that each Gd atom has the large positive magnetic moment value of 7.03 $\mu_B$, but the magnetic moments of Ti atoms at sites 3 and 4 in the III monolayer are close to zero because of the mutual cancellation of contributions from different occupied states. We show in Fig. 1(c) the charge density distribution from the energy window between -0.5 and 0 eV. It is clear that the major part comes from the Ti-O monolayer labelled with 'III' in the GTO layer.

\begin{table}[!h]
\caption{The average polar displacements along c direction of Ba, Ti, and Gd with the adjacent O in each monolayer in (BTO)$_m$/(GTO)$_2$ ($m$ = 4 and 2) superlattices. The positive and negative values implies that the cation Ba, Ti and Gd cations move leftward or rightward with respect to the O anion, respectively.}
\begin{ruledtabular}
\begin{tabular}{cccccc}
& layer & monolayer &   $m=4$       &    $m=2$  &  \\ \hline
& GTO  &  I       &   -0.12       &    -0.14   & \\
&     &  II      &  -0.14        &  -0.17      &\\
&     &  III       &  -0.07        &  -0.09      &\\
&     &  IV        &  -0.07        &  -0.08      &\\ \hline
& BTO  &  V        & -0.11, 0.40   &  -0.12, 0.38 &\\
&     &  VI       & 0.18          &  0.16       &\\
&     &  VII      & 0.05, 0.18    &  0.06, 0.13 &\\
&     &  VIII       & 0.10          &  0.10       &\\
&     &  IX        & 0.09, 0.11    &   - &\\
&     &  X       & 0.09          &   - &\\
&     &  XI      & 0.06, 0.14    &   - &\\
&     &  XII     & 0.07          &   - &
\end{tabular}
\end{ruledtabular}
\end{table}

Table I gives the Ba-O and Ti-O polar displacements along the $c$ axis. The black arrow in Fig. 1 indicates that the net polarization of the BTO layer points leftward. The largest polar displacement per monolayer, (-0.11 and 0.40 \AA), occurs at the interfacial Ti-O monolayer labelled with 'V' in Fig. 1, the next largest one (0.18 \AA) at the adjacent Ba-O layer labelled with 'VI', and the polar displacements of other Ti-O and Ba-O monolayers are near 0.10 \AA, which can weaken the polar discontinuity at the interfaces (IFA and IFB). Besides, we also found the polar displacement along the $c$ axis of Ti, Gd, and O in the GTO layer. In each monolayer, the Ti and Gd ions move rightward with respect to the neighboring O ions, which can reduce the diverging electrostatic potential in the GTO layer. The averaged displacements in each monolayer of the GTO layer are given in Table I. Because the ferroelectricity in BTO is originated from ionic displacements, these polar displacements, with the latest experimental results in high-quality ultra-thin BTO layers\cite{30a}, make us believe that the ferroelectricity truly exists in the BTO layer, and coexists with the conducting layers in these BTO/GTO superlattices. The ferroelectric polarization in the BTO layer can produce an electric field in the GTO layer, hindering the electronic reconstruction like that in LTO/STO and STO/GTO. Therefore, in the (BTO)$_4$/(GTO)$_2$ superlattice, there is only small amount of electrons per Ti atom in the interfacial Ti-O monolayer labelled with 'V' (including the Ti atoms labelled with '5' and '6') in Fig. 1(c), and these electrons populate the majority-spin d$_{xy}$ states, resulting in the d$_{xy}$ states splitting off the  d$_{zx}$ and  d$_{yz}$ states, as shown in Fig. 2, and a small magnetic moments of $0.09 \mu_B$ per Ti atom in this monolayer.

\subsection{Monolayer-resolved electronic structures}

\begin{figure}[!htbp]
\centering  
\includegraphics[clip, width=8.5cm]{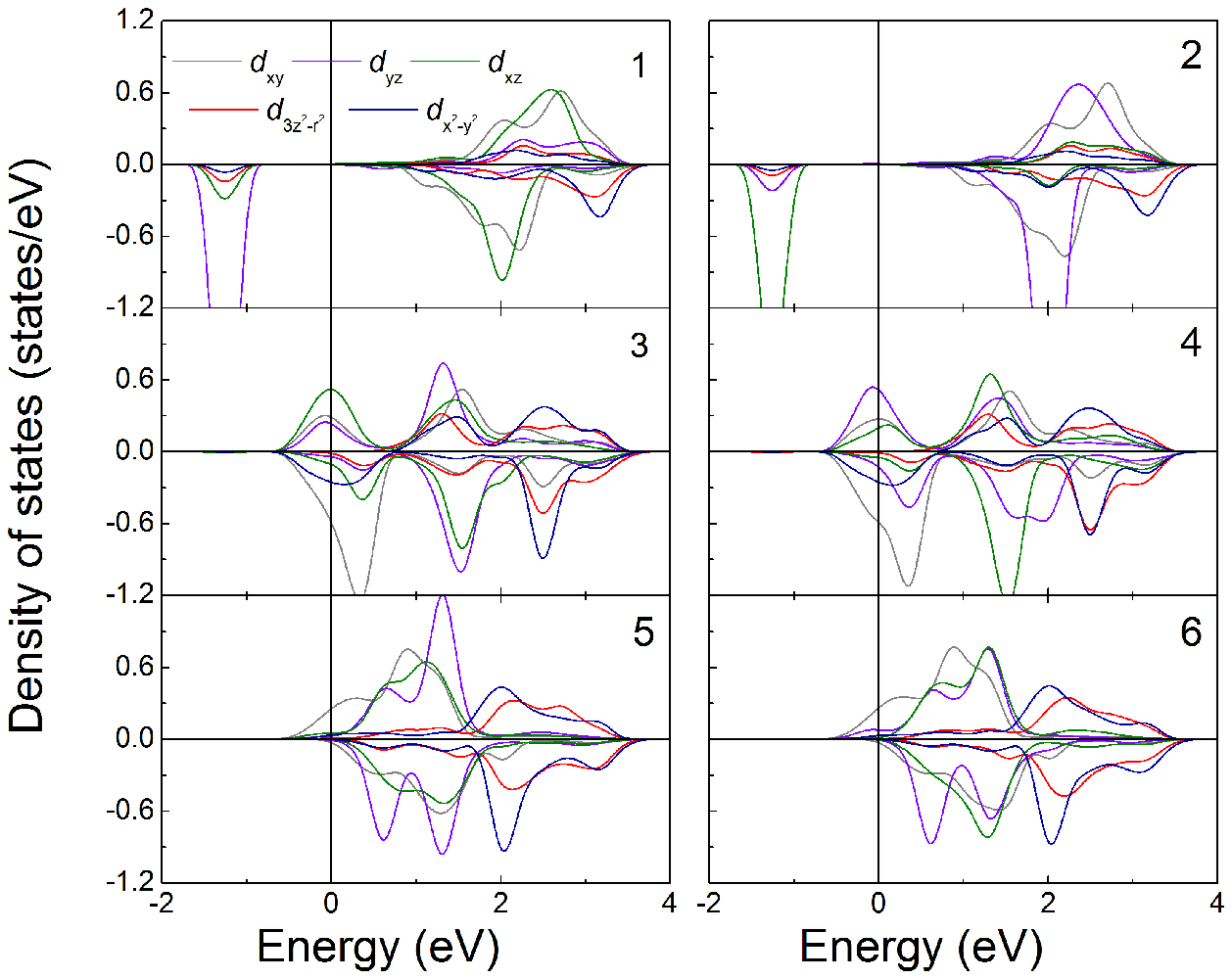}
\caption{(Color online) Orbital-resolved DOS for Ti atoms in the Ti-O monolayers labelled with 'I' (1 and 2), 'III' (3 and 4), and 'V' (5 and 6) in (BTO)$_4$/(GTO)$_2$ superlattice. }\label{edge}
\end{figure}
\begin{figure}[!htbp]
\centering  
\includegraphics[clip, width=8cm]{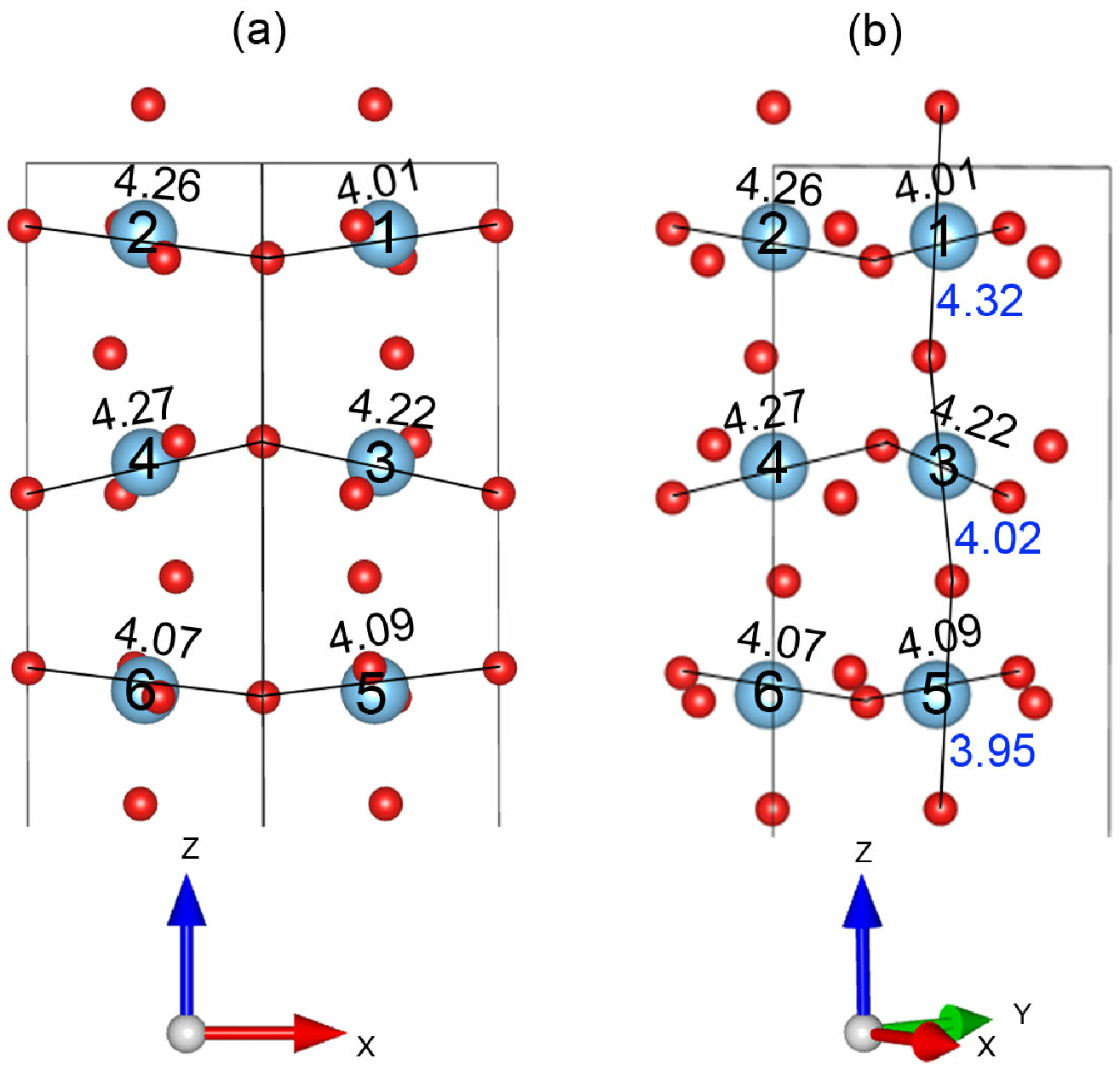}
\caption{(Color online) O-O distances of the three monolayers in the $ab$ plane and the $c$ axis of optimized Ti centered octahedra in (BTO)$_4$/(GTO)$_2$  in x-z plane (a), and rotated clockwise by 45$^\circ$ (b). The positions of Ti atoms are defined in Fig. 1(a).}\label{edge}
\end{figure}

\begin{figure*}[!tbp]
\centering  
\includegraphics[clip, width=14cm]{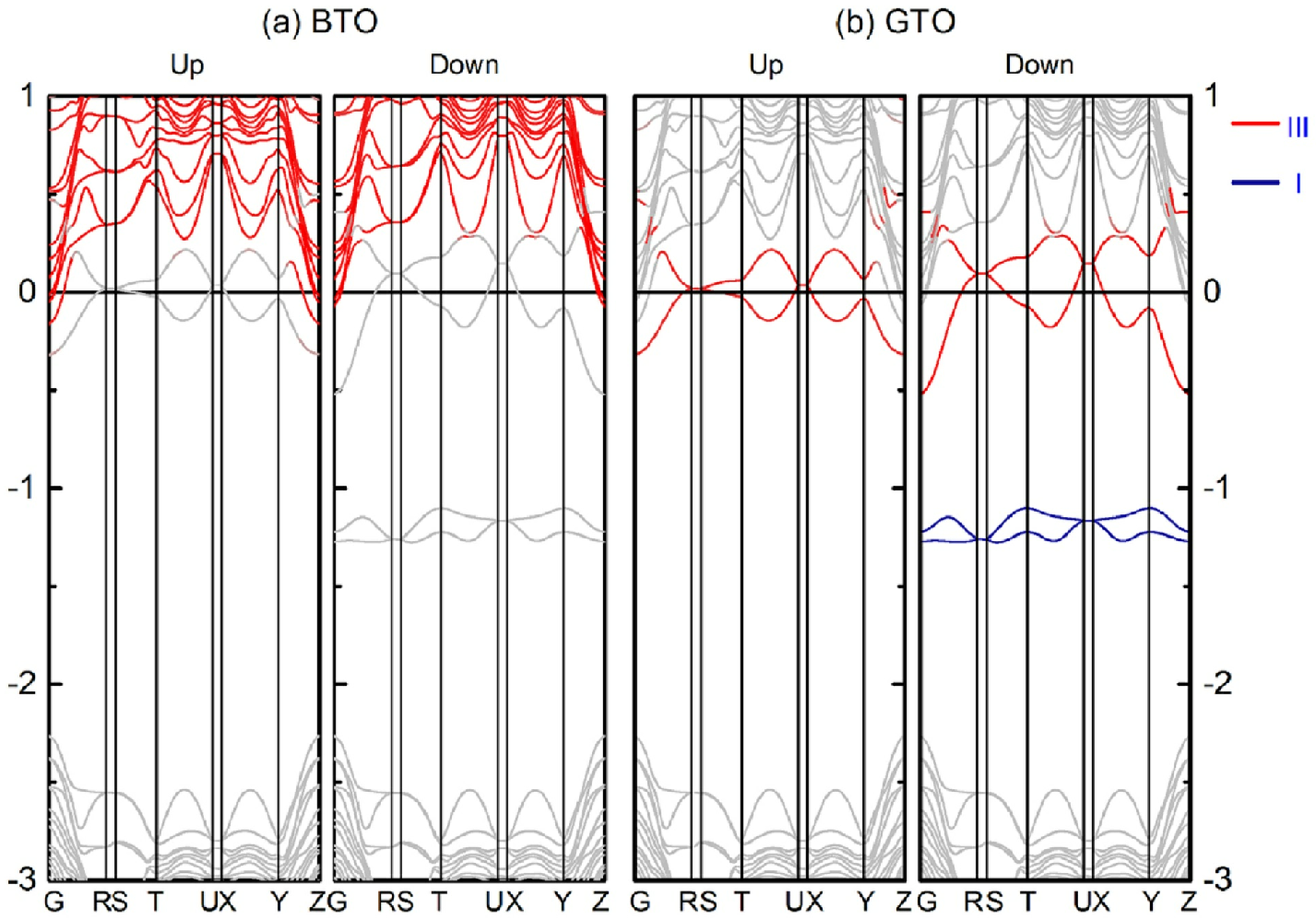}
\caption{(Color online) Spin-resolved band structure of the (BTO)$_4$/(GTO)$_2$ superlattice. (a) The red lines indicate the bands originated mainly from Ti atoms in the BTO layer, but the filled part of them is from the interfacial Ti-O monolayer labelled with 'V'. (b) The royal and red lines denote the bands of the Ti atoms at the '1' and '2' sites in the I monolayer and the bands near the Fermi level from the '3' and '4' Ti atoms in the III monolayer, respectively. }\label{edge}
\end{figure*}

In Fig. 2 we also present the orbital-resolved density of states (DOS) of Ti atoms in the Ti-O monolayers labelled with 'I' and 'III' in (BTO)$_4$/(GTO)$_2$ superlattice. It is clear that in the two monolayers, the d$_{yz}$ and d$_{zx}$ orbitals appear alternately at every second Ti atom in each of the Ti-O monolayers. This type of orbital ordering was observed experimentally in the ferromagnetic YTiO$_3$ through NMR \cite{25,26}, polarized neutron diffraction\cite{27}, and resonant x-ray scattering\cite{28}. In Fig. 3, we present the O-O distances over the bridging Ti atom for the three Ti-O monolayers. The O-O distance of 3.95 \AA{} (near the Ti atom labelled with '5') along the $z$ axis, shorter than those of 4.09 and 4.07 \AA{} along the $x$ and $y$ axes, causes the d$_{xy}$ orbital to have the lowest energy level for these two Ti atoms. The TiO$_6$ octahedron in the GTO layer is elongated along the $y$ direction at sites 1 and 3, while  they are elongated along the $x$ direction at sites 2 and 4, which can be seen by comparing Fig. 3(a) and 3(b). At sites 1 and 2, the occupied minority-spin states are mainly d$_{yz}$ or d$_{xz}$  orbital because the much longer O-O distances of 4.26 and 4.32 \AA{} along the $y$ (or $x$) and $z$ direction than the O-O distance of 4.01 \AA{} along the $x$ (or $y$) reduces the repulsive potential from the surrounding O ions in these two directions. These two Ti atoms contribute -0.88 and -0.89 $\mu_B$ to the magnetic moment, respectively, and there is an energy gap of 2.0 eV across the Fermi level. These magnetic moment values are consistent with the bulk value of -0.89 $\mu_B$.
The occupation of the majority-spin  d$_{xz}$,  d$_{xy}$, and d$_{yz}$   and the minority-spin d$_{xy}$  and d$_{x^2-y^2}$  result in the net magnetic moments of 0.01 and -0.03 $\mu_B$ for the Ti atoms at sites 3 and 4, respectively. From Fig. 3(b), it can be seen that the O-O distances in the $y$ (or $x$) and $x$ (or $y$) directions are 4.27 and 4.22 \AA, respectively, and the distance difference between the two directions is 0.05 \AA. It should be pointed out that in this Ti centered octahedron, the O-O distance in the $z$ direction is shortened by 0.25 and 0.20 \AA{} compared to those in the $y$ (or $x$) and $x$ (or $y$) directions. As a result, there are no occupied  d$_{3z^2-r^2}$ states for these two Ti atoms because of their high energy levels.

Fig. 4 displays the spin-resolved band structure of the (BTO)$_4$/(GTO)$_2$ superlattice. It can be seen that the system shows metallic property. The filled parts of the red bands in Fig. 4(a) are mainly originated from the Ti atoms in the V monolayer (near one interface), and the red and royal bands in Fig. 4(b) mainly from the Ti atoms of the III monolayer and the I monolayer (near the other interface). It is clear that the bands at the Fermi level are mainly from the Ti atoms at sites 3 and 4 in the III Ti-O monolayer, and a little part of the bands from the Ti atoms in the V monolayer. Therefore, the metallic feature of this system is mainly from the conductive Ti-O monolayer within the GTO layer, plus a little from one of the interfacial Ti-O monolayers, which is visually shown in Fig. 1(c). This phenomenon is interesting for spintronics applications because of the special 2DEG in the GTO layer. Our calculated results show that these 2DEG systems are almost the same for $m$= 4 and 2. It can be reasonably believed that the main 2DEG feature should remain the same when the parameter $m$ becomes larger than 4.

\subsection{Further insight}

In bulk GTO, the FM Ti array is antiparallel to the FM Gd array and the AFM Gd-Ti interactions were reported to be weaker than the FM Ti-Ti interactions.\cite{3,4} The spin-exchange along the $c$ axis in the perovskite Ti oxide $R$TiO$_3$ (where $R$ represents the trivalent rare-earth ions) changes from AFM to FM with increasing GdFeO$_3$-type distortion, with the transition occurring at GTO.\cite{29} For sites 1 and 2 in the 'I' monolayer, the orbitals of neighboring Ti atoms in the $ab$-plane are approximately orthogonal to each other. Hence, the FM spin configuration is favored through the Hund's-rule coupling. The various GdFeO$_3$-type distortions for Ti centered octahedral leads to the FM or AFM spin arrangement along the $c$ axis for Ti atoms. In Table II, we present the Ti-O bond lengthes and Ti-O-Ti bond angles in the I, III, and V monolayers. For the I monolayer, the Ti-O bond lengthes are so anisotropic that the oxygen octahedron is substantially distorted, causing the d$_{xz}$ or d$_{yz}$ split from the other d orbitals. For the III and V monolayers, the distorted oxygen octahedra are also consistent with the electronic structure feature. The Ti-O-Ti bond angle of 144.1$^\circ$, across the I, II, and III monolayers, is larger than that of bulk value of 139.5$^\circ$. This means the magnitude of the GdFeO$_3$-type distortion of the GTO layer is weaker than that in bulk GTO due to the tensile strain of 3.77\% in the $a$ axis for the GTO layer and the rearrangement of ionic positions, resulting in spin exchange along the $c$ axis being AFM. For Ti atoms in the Ti-O monolayer labelled with 'III' in the GTO layer, the coordination number of Ti with d$^1$ reduces by half, but that of Gd in the adjacent Gd-O monolayers is not changed. Therefore, for these Ti atoms, the co-existing antiparallel Gd-Ti and Ti-Ti couplings lead to the occupation of different orbitals in different spin channels in the same energy range near the fermi level.

\begin{table}[!h]
\caption{The Ti-O bond lengths ($l_{p}$, $l_z$ in \AA) and the Ti-O-Ti bond angles ($\theta_p$, $\theta_z$ in degree) in the ab plane and along the c axis for the three Ti-O monolayers (I, III, V) related to the GTO layer of (BTO)$_4$/(GTO)$_2$ superlattice.}
\begin{ruledtabular}
\begin{tabular}{cccccccc}
    	     &  GTO bulk   &              &    superlattice   &         \\ \hline
             &      -       &     I       &         III        &     V     \\ \hline
$l_{p}$ &  2.14, 2.06 &  2.09, 2.18  &    2.13, 2.14     &  2.03, 2.06 \\
             &             &  1.96, 2.06  &    2.11, 2.12     &  1.92, 2.18 \\
$l_z$  &  2.08       &  2.22        &    1.98           &  2.02  \\
             &             &  2.13        &    2.05           &  1.96  \\ \hline
$\theta_p$ & 143.3   &  151.5, 155.4 &  143.0, 143.1    &  162.2, 158.1 \\
$\theta_z$ & 139.5   & 172.9, 144.1 & 144.1, 147.1      & 147.1, 171.5
\end{tabular}
\end{ruledtabular}
\end{table}

\begin{figure}[!htbp]
\centering  
\includegraphics[clip, width=8cm]{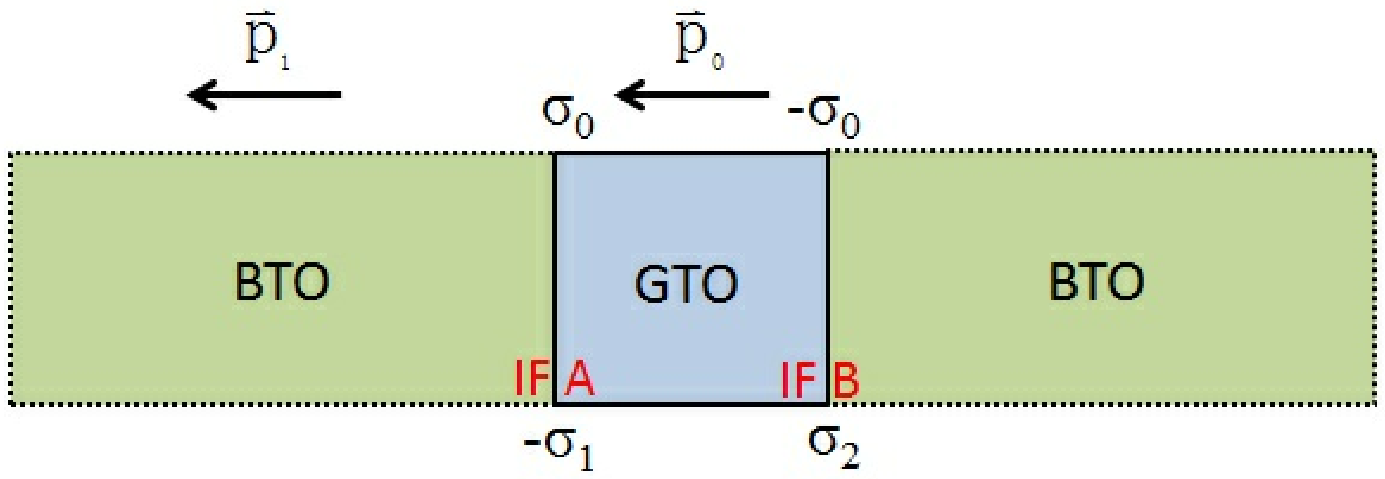}
\caption{(Color online) A demonstration of the polarization vectors ($\vec{P}_0$ and $\vec{P}_1$) of the GTO and BTO layers, and the virtual interfacial charges ($\pm\sigma_0$, $-\sigma_1$, and $\sigma_2$) at the two interfaces (IF A and IF B) in the BTO/GTO superlattices. }\label{edge}
\end{figure}

It can be seen from the band structure and the density of states that the two interfacial Ti-O monolayers are very different from each other. The I monolayer has one d electron per Ti atom and thus its Ti atom assumes +4 valence, but The Ti atom in the V monolayer approximately has +3 valence. This implies that there is no electron reconstruction between the interfaces in this case. It is in contrast to the STO/LTO and STO/GTO interfaces where an electron reconstruction happens, meaning that 1/2 electron per Ti atom is transferred from one interfacial Ti-O monolayer to the other.\cite{8,9,10,11,12,13,14} Without electron reconstruction, the LTO and GTO layers are polar in the z axis, and after the electron reconstruction, they become non-polar in the z axis. In our case, the GTO layer remains polar because there is no electron reconstruction! This result can be explained in terms of our calculated results by using Fig. 5. The polar GTO layer has a polarization vector $\vec{P}_0$ and it induces virtual interfacial charges $\pm\sigma_0$ at the two interfaces (IFA and IFB). The ferroelectric BTO layer has a polarization vector $\vec{P}_1$ which is parallel to $\vec{P}_0$. $\vec{P}_1$ induces the two interfacial charges, $-\sigma_1$ and $\sigma_2$, at the two interfaces, but $-\sigma_1$ and $\sigma_2$ are opposite to the interfacial charges from $\vec{P}_0$. Therefore, the interfacial charges and the intrinsic electric field in the GTO layer are substantially weakened due to the ferroelectric polarization, which avoids electron reconstruction between the two interfacial Ti-O monolayers and thus makes the asymmetrical 2DEG and should achieve interesting Rashba spin-orbit effects.

\section{Conclusion}

In summary, we have studied crystal structures, electronic states, and magnetism of short-period superlattices consisting of ferroelectric BTO and ferrimagnetic insulating polar GTO in terms of first-principles calculations. We have optimized all the cell volumes and atomic positions of (BTO)$_m$/(GTO)$_2$ ($m$=2 and 4) with GGA+U method. Our investigation shows that the ferroelectricity in the insulating BTO layer induces an inhomogeneous electric field against the polarity-produced electric field in the GTO layer and thus differentially changes the d energy levels of the three Ti-O monolayers related with the GTO layer. Consequently, these make the middle Ti-O monolayer in the GTO layer become conductive. Through avoiding electron reconstruction between the two interfacial Ti-O monolayers, the ferroelectric polarization also makes the electronic states and magnetism of interfacial Ti-O monolayers become substantially different from those in the GTO/STO superlattices without ferroelectricity. These results make us believe that these main electronic properties related with the GTO layer should be almost the same for larger $m$. Therefore, these 2DEG systems are interesting for potential spintronics applications because of their unique asymmetrical two-dimensional properties and possible useful spin-orbit effects.

\begin{acknowledgments}
This work is supported by the Nature Science Foundation of China (Grant No. 11174359 and No. 11574366), by the Department of Science and Technology of China (Grant No. 2016YFA0300701 and No. 2012CB932302), and by the Strategic Priority Research Program of the Chinese Academy of Sciences (Grant No.XDB07000000).
\end{acknowledgments}

\end{document}